\documentclass[runningheads]{llncs}
\usepackage{graphicx}
\usepackage{tabularx}
\usepackage{color}
\usepackage{todonotes}
\begin{document}
\title{HateMonitors: Language Agnostic Abuse Detection in Social Media}
\author{Punyajoy Saha\inst{1}\orcidID{0000-0002-3952-2514} \and
Binny Mathew\inst{1}\orcidID{0000-0003-4853-0345} \and
Pawan Goyal\inst{1}\orcidID{0000-0002-9414-8166} \and
Animesh Mukherjee\inst{1}\orcidID{0000-0003-4534-0044}
}
\authorrunning{P. Saha et al.}

\institute{Indian Institute of Technology, Kharagpur, West Bengal, India - 721302}
\maketitle              % typeset the header of the contribution
\begin{abstract}
Reducing hateful and offensive content in online social media pose a dual problem for the moderators. On the one hand, rigid censorship on social media cannot be imposed. On the other, the free flow of such content cannot be allowed. Hence, we require efficient abusive language detection system to detect such harmful content in social media. In this paper, we present our machine learning model, HateMonitor, developed for Hate Speech and Offensive Content Identification in Indo-European Languages (HASOC)~\cite{hasoc2019overview}, a shared task at FIRE 2019. We have used Gradient Boosting model, along with BERT and LASER embeddings, to make the system language agnostic. Our model came at \textbf{First position} for the German sub-task A. We have also made our model public~\footnote{\url{https://github.com/punyajoy/HateMonitors-HASOC}}.

\end{abstract}
\section{Introduction}

In social media, abusive language denotes a text which contains any form of unacceptable language in a post or a comment. Abusive language can be divided into hate speech, offensive language and profanity. Hate speech is a derogatory comment that hurts an entire group in terms of ethnicity, race or gender. Offensive language is similar to derogatory comment, but it is targeted towards an individual. Profanity refers to any use of unacceptable language without a specific target. While profanity is the least threatening, hate speech has the most detrimental effect on the society.

Social media moderators are having a hard time in combating the rampant spread of hate speech\footnote{\url{https://tinyurl.com/y6tgv865}} as it is closely related to the other forms of abusive language. The evolution of new slangs and multilingualism, further adding to the complexity. 

Recently, there has been a sharp rise in hate speech related incidents in India, the lynchings being the clear indication \cite{arun2019whatsapp}. Arun et al.~\cite{arun2019whatsapp} suggests that hate speech in India is very complicated as people are not directly spreading hate but are spreading misinformation against a particular community. Hence, it has become imperative to study hate speech in Indian language.

For the first time, a shared task on abusive content detection has been released for Hindi language at HASOC 2019. This will fuel the hate speech and offensive language research for Indian languages. The inclusion of datasets for English and German language will give a performance comparison for detection of abusive content in high and low resource language.

In this paper, we focus on the detection of multilingual hate speech detection that are written in Hindi, English, and German and describe our submission \textit{(HateMonitors)} for HASOC at FIRE 2019 competition. Our system concatenates two types of sentence embeddings to represent each tweet and use machine learning models for classification.

\section{Related works}

Analyzing abusive language in social media is a daunting task. Waseem et al.~\cite{waseem2017understanding} categorizes abusive language into two sub-classes -- hate speech and offensive language. In their analysis of abusive language,  Classifying abusive language into these two subtypes is more challenging due to the correlation between offensive language and hate speech~\cite{davidson2017automated}. Nobata et al.~\cite{nobata2016abusive} uses predefined language element and embeddings to train a regression model. With the introduction of better classification models \cite{qian2018hierarchical,stammbach2018offensive} and newer features \cite{alorainy2018enemy,davidson2017automated,unsvaag2018effects}, the research in hate and offensive speech detection has gained momentum.

Silva et al. ~\cite{silva2016analyzing} performed a large scale study to understand the target of such hate speech on two social media platforms: Twitter and Whisper. These target could be the Refugees and Immigrants~\cite{ross2017measuring}, Jews~\cite{bilewicz2013harmful,finkelstein2018quantitative} and Muslims~\cite{awan2016islamophobia,vidgen2018detecting}. People could become the target of hate speech based on Nationality~\cite{erjavec2012you}, sex~\cite{bartlett2014misogyny,saha2018hateminers}, and gender~\cite{reddy2002perverts,gatehouse2018troubling} as well. Public expressions of hate speech affects the devaluation of minority members~\cite{greenberg1985effect}, the exclusion of minorities from the society~\cite{mullen2003ethnophaulisms}, and tend to diffuse through the network at a faster rate~\cite{mathew2019spread}.

One of the key issues with the current state of the hate and offensive language research is that the majority of the research is dedicated to the English language on~\cite{fortuna2018survey}. Few researchers have tried to solve the problem of abusive language in other languages \cite{ross2017measuring,sanguinetti2018italian}, but the works are mostly monolingual. Any online social media platform contains people of different ethnicity, which results in the spread of information in multiple languages. Hence, a robust classifier is needed, which can deal with abusive language in the multilingual domain. Several shared tasks like HASOC~\cite{hasoc2019overview}, HaSpeeDe~\cite{bosco2018overview}, GermEval~\cite{wiegand2018overview}, AMI~\cite{fersini2018overview}, HatEval~\cite{basile2019semeval} have focused on detection of abusive text in multiple languages recently.

\section{Dataset and Task description}

The dataset at HASOC 2019 \footnote{\url{https://hasoc2019.github.io/}} were given in three languages: Hindi, English, and German. Dataset in Hindi and English had three subtasks each, while German had only two subtasks.  We participated in all the tasks provided by the organisers and decided to develop a single model that would be language agnostic. We used the same model architecture for all the three languages.

\subsection{Datasets}
We present the statistics for HASOC dataset in Table~\ref{tab:dataset_statistics}. From the table, we can observe that the dataset for the German language is highly unbalanced, English and Hindi are more or less balanced for sub-task A. For sub-task B German dataset is balanced but others are unbalanced. For sub-task C both the datasets are highly unbalanced.

\begin{table}[!htbp]
\centering
\caption{This table shows the initial statistics about the training and test data}\label{tab:dataset_statistics}
\begin{tabular}{|c|c c||c c||c c|}
\hline
Language &  \multicolumn{2}{c||}{English} & \multicolumn{2}{c||}{German} & \multicolumn{2}{c|}{Hindi}\\
\hline
Sub-Task A & Train & Test & Train & Test & Train & Test  \\
\hline
HOF & 2261 & 288 & 407 & 136 & 2469 & 605\\
NOT & 3591 & 865 &3142 & 714 & 2196 & 713\\
\hline
Total & 5852 &1153& 3819 & 850 & 4665 & 1318\\
\hline \hline
Sub-Task B & Train & Test & Train & Test & Train & Test \\
\hline
HATE & 1141 &124 & 111 & 41 & 556 &190 \\
OFFN & 451 & 71 & 210 & 77 & 676 & 197\\
PRFN & 667 & 93 & 86 & 18 & 1237 & 218\\
\hline
Total & 2261 & 288 & 407 & 136 & 2469 & 605\\
\hline \hline
Sub-Task C & Train & Test & Train & Test & Train & Test \\
\hline
TIN & 2041 & 245 & - - & - - & 1545 & 542\\
UNT & 220 & 43 & - - & - - & 924 & 63\\
\hline
Total & 2261 & 288 & - - & - - & 2469 & 605\\
\hline
\end{tabular}
\end{table}

\subsection{Tasks}
% There were at most three subtask for each language which are enlisted below.

\textbf{Sub-task A}
consists of building a binary classification model which can predict if a given piece of text is hateful and offensive (HOF) or not (NOT). 
A data point is annotated as HOF if it contains any form of non-acceptable language such as hate speech, aggression, profanity. Each of the three languages had this subtask.

\noindent\textbf{Sub-task B}
consists of building a multi-class classification model which can predict the three different classes in the data points annotated as HOF: Hate speech (HATE), Offensive language (OFFN), and Profane (PRFN). Again all three languages have this sub-task.

\noindent\textbf{Sub-task C}
consists of building a binary classification model which can predict the type of offense: Targeted (TIN) and Untargeted (UNT).
Sub-task C was not conducted for the German dataset.

\section{System Description}

\begin{figure}
 \centering
 \includegraphics[width=0.6\textwidth]{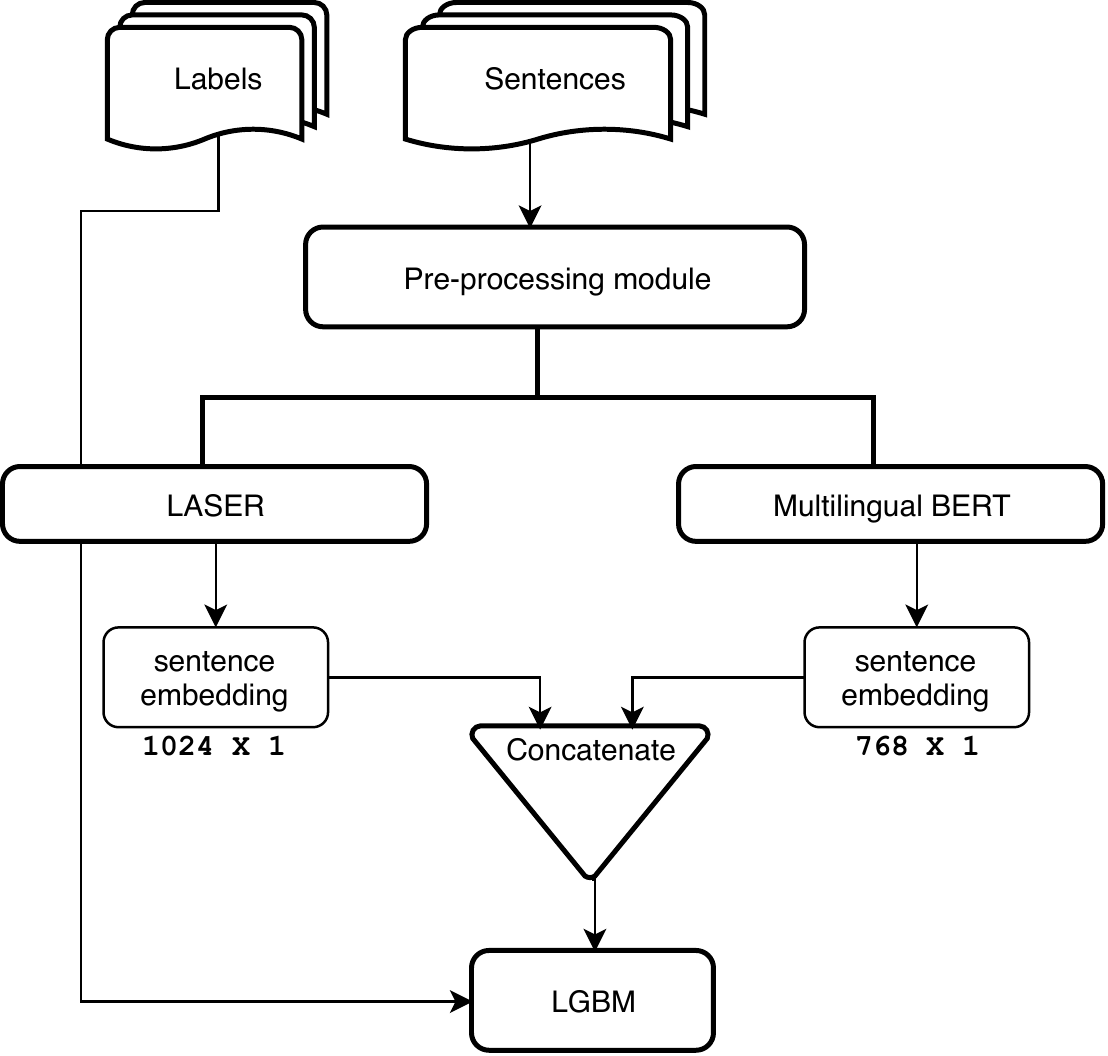}
 \caption{Architecture of our system} \label{Train}
 \end{figure}

In this section, we will explain the details about our system, which comprises of two sub-parts- feature generation and model selection. Figure \ref{Train} shows the architecture of our system.

\subsection{Feature Generation}

\subsubsection{Preprocessing:}
We preprocess the tweets before performing the feature extraction. The following steps were followed:

\begin{itemize}
    \item We remove all the URLs.
    
    \item Convert text to lowercase. This step was not applied to the Hindi language since Devanagari script does not have lowercase and uppercase characters.
    \item We did not normalize the mentions in the text as they could potentially reveal important information for the embeddings encoders. 
    \item Any numerical figure was normalized to a string `number'.
    
\end{itemize}

We did not remove any punctuation and stop-words since the context of the sentence might get lost in such a process. Since we are using sentence embedding, it is essential to keep the context of the sentence intact.

\subsubsection{Feature vectors:}
The preprocessed posts are then used to generate features for the classifier. For our model, we decided to generate two types of feature vector: BERT Embeddings and LASER Embeddings. For each post, we generate the BERT and LASER Embedding, which are then concatenated and fed as input to the final classifier.

\textbf{Multilingual BERT embeddings:} Bidirectional Encoder Representations from Transformers(BERT) \cite{DBLP:journals/corr/abs-1810-04805} has played a key role in the advancement of natural language processing domain (NLP). BERT is a language model which is trained to predict the masked words in a sentence. To generate the sentence embedding\footnote{We use the BERT-base-multilingual-cased which has 104 languages, 12-layer, 768-hidden, 12-heads and 110M parameters} for a post, we take the mean of the last 11 layers (out of 12) to get a sentence vector with length of 768.

\textbf{LASER embeddings}: Researchers at Facebook released a language agnostic sentence embeddings representations (LASER) \cite{DBLP:journals/corr/abs-1812-10464}, where the model jointly learns on 93 languages. The model takes the sentence as input and produces a vector representation of length 1024. The model is able to handle code mixing as well~\cite{VERMA1976153}.

We pass the preprocessed sentences through each of these embedding models and got two separate sentence representation. Further, we concatenate the embeddings into one single feature vector of length 1792, which is then passed to the final classification model.

\subsection{Our Model}
The amount of data in each category was insufficient to train a deep learning model. Building such deep models would lead to overfitting. So, we resorted to using simpler models such as SVM and Gradient boosted trees. Gradient boosted trees~\cite{DBLP:journals/corr/ChenG16} are often the choice for systems where features are pre-extracted from the raw data\footnote{\url{https://tinyurl.com/yxmuwzla}}. In the category of gradient boosted trees, Light Gradient Boosting Machine (LGBM) \cite{Ke2017LightGBMAH} is considered one of the most efficient in terms of memory footprint. Moreover, it has been part of winning solutions of many competition \footnote{\url{https://tinyurl.com/y2g8nuuo}}. Hence, we used LGBM as model for the downstream tasks in this competition.

\section{Results}
The performance of our models across different languages for sub-task A are shown in table \ref{tab2}. Our model got the \textbf{first position} in the German sub-task with a macro F1 score of \textbf{0.62}. The results of sub-task B and sub-task C is shown in table \ref{tab3} and \ref{tab4} respectively.

\begin{table}[!tbp]
\centering
\caption{This table gives the language wise result of sub-task A by comparing the macro F1 values}\label{tab2}
\begin{tabular}{|c|c|c|c|}
\hline
Language &  English & German & Hindi\\
\hline
HOF & 0.59 & 0.36  & 0.76 \\
\hline
NOT & 0.79 & 0.87 & 0.79\\
\hline
Total & 0.69 & 0.62 &0.78\\
\hline
\end{tabular}
\end{table}

\begin{table}[!tbp]
\centering
\caption{This table gives the language wise result of sub-task B by comparing the macro F1 values }\label{tab3}
\begin{tabular}{|c|c|c|c|}
\hline
Language &  English & German & Hindi\\
\hline
HATE & 0.28 & 0.04  & 0.29 \\
\hline
OFFN & 0.00 & 0.0 & 0.29\\
\hline
PRFN & 0.31 & 0.19 & 0.59\\
\hline
NONE & 0.79 & 0.87 & 0.79\\
\hline
Total & 0.34 & 0.28 &0.49\\
\hline
\end{tabular}
\end{table}

\begin{table}[!tbp]
\centering
\caption{This table gives the language wise result of sub-task C by comparing the macro F1 values}\label{tab4}
\begin{tabular}{|c|c|c|}
\hline
Language &  English & Hindi\\
\hline
TIN & 0.51 & 0.63 \\
\hline
UNT & 0.11 & 0.17\\
\hline
NONE & 0.79 & 0.79\\
\hline

Total & 0.47 & 0.53\\
\hline
\end{tabular}
\vspace{-0.5cm}
\end{table}

\section{Discussion}

In the results of subtask A, models are mainly affected by imbalance of the dataset. The training dataset of Hindi dataset was more balanced than English or German dataset. Hence, the results were around \textbf{0.78}. As the dataset in German language was highly imbalanced, the results drops to \textbf{0.62}. In subtask B, the highest F1 score reached was by the profane class for each language in table \ref{tab3}. The model got confused between OFFN, HATE and PRFN labels which suggests that these models are not able to capture the context in the sentence. The subtask C was again a case of imbalanced dataset as targeted(TIN) label gets the highest F1 score in table \ref{tab4}.

\section{Conclusion}

In this shared task, we experimented with zero-shot transfer learning on abusive text detection with pre-trained BERT and LASER sentence embeddings. We use an LGBM model to train the embeddings to perform downstream task. Our model for German language got the first position. The results provided a strong baseline for further research in multilingual hate speech. We have also made the models public for use by other researchers\footnote{\url{https://github.com/punyajoy/HateMonitors-HASOC}}.

\bibliographystyle{splncs04}
\bibliography{main}
\end{document}